\documentclass[twocolumn,superscriptaddress,showpacs,nofootinbib,preprintnumbers,secnumarabic,amssymb, nobibnotes, aps, prd]{revtex4-2}
\usepackage[utf8]{inputenc}
\usepackage{graphicx}
\usepackage{latexsym,amsmath,amssymb,amsthm,lmodern,float,url}
\usepackage{natbib}
\usepackage{color}
\usepackage{microtype}
\usepackage{import}
\usepackage{bbold}
\usepackage[plain]{fancyref}
\usepackage{varioref}
\usepackage{slashed}
\usepackage{multirow}
\usepackage{tikz}
\usepackage{scrextend}
\usepackage{braket}
\usepackage{units}
\usetikzlibrary{shapes}
\usetikzlibrary{positioning}
\usepackage[normalem]{ulem}

\newcommand{\fig}[1]{Fig.~\ref{fig:#1}}

\newcommand{\eq}[1]{Eq.~(\ref{eq:#1})}
\newcommand{\Sec}[1]{Sec.~(\ref{sec:#1})}
\newcommand*{\dif}{\mathop{}\!\mathrm{d}}

\usepackage[colorlinks=true,backref=false, linktocpage=true,
citecolor=blue,urlcolor=blue,linkcolor=blue,pdfpagemode=UseOutlines]{hyperref}
\hypersetup{%
  bookmarksnumbered=true,
  pdftitle = {},
  pdfsubject = {},
  pdfauthor = {},
  pdfkeywords = {}
}

\begin{document}
\preprint{\hspace{0.15cm}USTC-ICTS/PCFT-24-17}
\title{Neutrino magnetic dipole portal with low energy neutrino nucleus scattering data}
\author{Ying-Ying Li}
\email{yingyingli@ustc.edu.cn}
\affiliation{Interdisciplinary Center for Theoretical Study, University of Science and Technology of China, Hefei, Anhui 230026, China}
\affiliation{Peng Huanwu Center for Fundamental Theory, Hefei, Anhui 230026, China}
\author{Yu-Feng Li}
\email{liyufeng@ihep.ac.cn}
\affiliation{Institute of High Energy Physics, Chinese Academy of Sciences, Beijing 100049, China}
\affiliation{School of Physical Sciences, University of Chinese Academy of Sciences, Beijing 100049, China}
\author{Shuo-Yu Xia}
\email{xiashuoyu@ihep.ac.cn}
\affiliation{Institute of High Energy Physics, Chinese Academy of Sciences, Beijing 100049, China}
\affiliation{School of Physical Sciences, University of Chinese Academy of Sciences, Beijing 100049, China}
\date{\today}

\begin{abstract}
Sterile neutrinos that couple to the Standard Model via the neutrino magnetic dipole portals have been extensively studied at various experiments. In this work, we scrutinize these interactions for sterile neutrinos in the mass range of $\unit[0.1]{}-\unit[50]{MeV}$ 
through the nuclear and electron recoils at various neutrino scattering experiments. For the $e$-flavor specific dipole portal, we demonstrate that Dresden-II can provide leading constraints for $m_N \lesssim \unit[0.5]{MeV}$, setting aside currently unresolved theoretical uncertainties. For the $\mu$-flavor case, we show that the COHERENT experiment can probe a unique parameter region for $m_N$ in the range of $\unit[10]{}-\unit[40]{MeV}$ with the full dataset collected by the CsI[Na]
scintillation detector, including both the energy and timing structure of the neutrino beam. We also present limits on the parameter regions of the $\tau$-flavor dipole portal using measurements of the solar neutrino flux from dark matter direct detection experiments. 

\end{abstract}

\maketitle

\section{Introduction}
Observations of non-zero neutrino mass point to physics beyond the Standard Model (BSM), which generically extends the neutrino sector with new singlet fermionic states. These new states, dubbed as sterile neutrinos, can couple to the Standard Model (SM) sector via its mixing with active neutrinos at the renormalizable level (see e.g. review \cite{Dasgupta_2021}). Sterile neutrino can also couple to the SM sector via the a dimension-five neutrino dipole operator (NDP), the Lagrangian of which after electroweak symmetry breaking can be written as~\cite{Schwetz:2020xra,Magill:2018jla}
	\begin{equation}
	\mathcal{L}_{\rm NDP}=\frac{d_\alpha}{2}(\bar{N}\sigma_{\mu\nu}\nu^\alpha F^{\mu\nu})+ {\rm h.c.}
	\label{lagrangian}	\end{equation}
where $\nu^\alpha$ is the active neutrino of flavor $\alpha$, $N$ is the sterile neutrino and $d_\alpha$ is the neutrino transitional magnetic moment. Without specifying the flavor, we will denote $d_\alpha$ as $d$. In the last few years, significant attentions have been paid to search for $N$ via this dipole portal. In the high sterile neutrino mass $m_N$ regime, LEP and LHC can probe the magnetic moment at the level of $d\sim 1/(\unit[10^3]{GeV})$ \cite{Magill_2018}. In the sub-GeV mass region, sterile neutrinos produced from Supernova explosion can escape and carry away energies. The energy loss requirements can exclude the band $ 1/(\unit[10^{7}]{GeV})\gtrsim d \gtrsim 1/(\unit[10^{11}]{GeV})$ \cite{Magill_2018}. Below this cooling band, looking for decay products of sterile neutrino originated from Supernova explosion at neutrino detectors \cite{Brdar:2023tmi,lazar2024supernovae} and $\gamma-$ray telescopes \cite{Brdar:2023tmi} can probe even smaller values of $d$.  

 To probe the parameter regions above the cooling band, high intensity beam dump and neutrino experiments are the main tools under consideration. There are searches via measuring neutrino scattering spectrum distortions at dark matter detectors~\cite{Shoemaker:2018vii, Brdar:2020quo}, upscattering of solar and atmospheric neutrinos to $N$ which decays at detectors \cite{Plestid:2020vqf, Gustafson:2022rsz}, Borexino $e-\nu$ scattering \cite{Brdar:2020quo, BOREXINO:2018ohr}, CHARM-II $e-\nu$ scattering \cite{Coloma:2017ppo, CHARM-II:1991ydz}, etc. With these efforts, parameter regions above the cooling band for $m_N \lesssim \unit[10]{MeV}$ can be comprehensively probed, while for larger $m_N$ up to $\unit[50]{MeV}$, existing studies can cover the parameter region to $d \gtrsim 1/(2\times\unit[10^{6}]{GeV})$, leaving an unexplored regions for smaller $d$.

On the other hand, the vast number of neutrinos produced from reactors and Spallation Neutron Source (SNS) is energetic enough to produce $N$ with $m_N \gtrsim \unit[]{MeV}$ via upscattering.  
Once happens, this process will modify the neutrino scattering data from the SM predictions, thus can be probed at experiments. Given its more energetic spectrum, upscattering of neutrinos from SNS can produce sterile neutrino as heave as $\unit[50]{MeV}$, which might cover the unexplored parameter regions mentioned above.
Until recently, coherent elastic neutrino-nucleus scattering (CE$\nu$NS) has been observed by COHERENT collaboration \cite{COHERENT:2017ipa}, and a suggestive evidence of CE$\nu$NS using reactor antineutrinos has also been reported by Dresden-II experiment \cite{Colaresi:2022obx}. Elastic neutrino-electron scattering (E$\nu$ES), a concurrent process to CE$\nu$NS, can contribute similarly to the scattering data and may not be separated from the nuclear recoil due to CE$\nu$NS. Though neutrino-electron scattering in the 
SM is negligible, it might obtain significant contributions from BSM. Studies in this direction appeared recently \cite{Dasgupta:2021fpn, Miranda:2021kre, Bolton:2021pey} where the measured energy spectra at COHERENT are used to constrain the neutrino magnetic dipole portals.  

In the work, we utilize Dresden-II experiment and the full dataset collected by the CsI[Na] detector at COHERENT experiment to probe sub-GeV sterile neutrinos. We will show that Dresden-II can probe a large parameter space of the $e$-flavor magnetic portal, while the full dataset together with the time information will enable COHERENT experiments to probe a unique parameter region for $N$ with $\unit[10]{MeV} \lesssim m_N \lesssim\unit[40]{MeV}$ for the $\mu$-flavor magnetic dipole portal. To constrain the $\tau$-flavor portal, we include the measurements of solar neutrino fluxes from PandaX-4T and Xenon1T experiment. We will present the theoretical framework to calculate the CE$\nu$NS and E$\nu$ES in \Sec{theory} for the SM and the NDP. In \Sec{spectra}, we discuss the energy spectra at various experiments for the neutrino magnetic dipole portal, and present the constraints obtained in \Sec{limts}. \Sec{conclusion} includes the conclusions and possible future directions to explore.

\section{Theoretical Framework}
\label{sec:theory}
In this section, we will briefly introduce the CE$\nu$NS and E$\nu$ES in the SM and the upscattering process from active neutrino to sterile neutrino based on NDP.
\subsection{CE$\nu$NS and E$\nu$ES in the Standard Model}
The differential cross section of the CE$\nu$NS process between neutrino with energy $E_{\nu}$ and a nucleus in the SM can be written as \cite{PhysRevD.30.2295, Barranco_2005, Patton_2012, Coloma:2022avw, AtzoriCorona:2022qrf}
\begin{equation}
\frac{\dif{\sigma^{\nu-\mathcal{N}}_{\mathrm{SM}}}}{\dif{T_{n}}}(E_{\nu},T_{n})=\frac{G_{F}^2  M}{\pi}\left(1-\frac{MT}{2E_{\nu}^2}\right)(Q_V^{\rm SM})^2\,,
\label{eq:csSM} 
\end{equation}
where $T_{n}$ is the kinetic energy of nuclear recoil, $M$ is the nucleus mass and $G_{F}$ is the Fermi constant. The cross section of CE$\nu$NS benefits from the coherent effect and is enhanced based on the square of the weak charge of the nucleus $Q_V^{\rm SM}$, which can be written as
\begin{equation}
Q_V^{\rm SM}=g_{V}^{p} Z F_{Z}\left(q^2\right) +g_{V}^{n} N F_{N}\left( \left|\vec{q}\right|  ^2\right)
\end{equation}
for a nucleus with $Z$ protons and $N$ neutrons. The vector neutrino-proton coupling $g^p_V$ and the vector neutrino-neutron coupling $g^n_V$ are given by
\begin{equation}
\begin{aligned}g_{V}^{p}(\nu_e)=0.0401,&\quad\quad g_{V}^{p}(\nu_\mu)=0.0318\,,\\
g_{V}^{p}(\nu_\tau)=0.0275,&\quad\quad g_{V}^{n}=-0.5094\,,
\end{aligned}
\end{equation}
where we employ the radiative corrections in the $\overline{\rm MS}$ scheme~\cite{Erler:2013xha} and derive the accurate values of the vector couplings.
$F_{Z}\left(q^2\right) $ and $ F_{N}\left(q^2\right) $ are the form factors of nucleon distributions in the nucleus for proton and neutron respectively and represent the Fourier transforms of the corresponding nucleon distribution in the nucleus. In this work, we employ the Helm parameterization~\cite{Helm:1956zz} with the Helm form factors depending on the the momentum transfer $\left|\vec{q}\right|$, the rms radii of neutron~\cite{Hoferichter:2020osn, Angeli:2013epw} and the rms radii of proton ~\cite{Fricke2004, Fricke:1995zz, Angeli:2013epw} in the nucleus. It should be noted that the cross section for CE$\nu$NS in the SM is universal for neutrinos of different flavors and the minimum neutrino energy to produce a recoil kinetic energy $T_{n}$ is $E_{\rm{min}}=\sqrt{MT_{n}/2}$ in the region $T_{n}\ll M$.

The differential cross section of the E$\nu$ES process between a neutrino and the electrons in the target can be written as \cite{Coloma:2022avw}
\begin{equation}
\begin{aligned}
\frac{\dif\sigma^{\nu-e}_{\rm SM}}{\dif T_e}(E,T_e)&=Z^{A}_{eff}(T_e)\frac{G^2_Fm_e}{2\pi}\Bigg[(g_V^{\nu_l}-g_A^{\nu_l})^2\left(1-\frac{T_e}{E} \right)^2\\
&+(g_V^{\nu_l}+g_A^{\nu_l})^2-((g_V^{\nu_l})^2-(g_A^{\nu_l})^2)\frac{m_eT_e}{E^2} \Bigg],
\end{aligned}
\end{equation}
where $T_e$ is the kinetic energy of electron recoil and $m_e = \unit[0.511]{MeV}$ is the electron mass. The cross section is not universal for electron neutrino and the neutrinos of the other two flavors since a electron neutrino can scatter with an electron through charged current process. This difference is represented with the flavor dependent neutrino electron couplings
\begin{equation}
\begin{aligned}
&g^{\nu_e}_V=2\sin^2\theta_W+1/2,\quad\quad g^{\nu_e}_A=1/2,\\
&g^{\nu_{\mu,\tau}}_V=2\sin^2\theta_W-1/2,\quad\quad g^{\nu_{\mu,\tau}}_A=-1/2.\\
\end{aligned}
\end{equation}
For antineutrinos, the vector couplings are given by $g^{\bar{\nu}_l}_V = g^{\nu_l}_V$ while the axial couplings satisfy the relation $g^{\bar{\nu}_{l}}_A=-g^{{\nu}_{l}}_A$. As the cross section is derived based on the Free Electron Approximation hypothesis, where the electrons are considered to be free and rest in the materials~\cite{Kouzakov:2017hbc,Kouzakov:2014lka,Chen:2014ypv,PhysRevD.100.073001}, the effective electron charge term $Z^{A}_{eff}(T_e)$ is employed to quantify the number of electrons that can be ionized with the deposited recoil kinetic energy $T_e$~\cite{Thompsonxray, Mikaelyan:2002nv,Fayans:2000ns}.

\subsection{Neutrino Dipole Portal}
The sterile neutrino, with mass $m_N$ can be produced via the active neutrino upscattering through the neutrino magnetic dipole portal. In this study, we will consider magnetic moment between active neutrino of a specific $\alpha$ and $N$ separately. For a general discussion in the following, we will write $d_\alpha$ as $d$, unless the flavor $\alpha$ is specific.
The cross section of producing $N$ via neutrino-nucleus scattering can be written as~\cite{Brdar:2020quo,Shoemaker:2018vii} 
\begin{equation}
 \begin{aligned}
\frac{\dif \sigma^{\nu-\mathcal{N}}_{\rm NDP}}{\dif T_n}&= d^2 \alpha_{\rm EM} Z^{2}F^{2}_{Z}(q^2)\bigg[-\frac{m_N^2}{2 E_{\nu} MT_n}\left(1-\frac{M-T_n}{2E_{\nu}}\right)\\
&-\frac{m_N^4(M-T_n)}{8E_{\nu}^{2}M^2T_n^2}+\frac{1}{T_n}-\frac{1}{E_{\nu}}\bigg],
 \end{aligned}
\label{eq:NDPxs}
\end{equation}
where $\alpha_{\rm EM}$ is fine structure constant. Contributions to \eq{NDPxs} from nuclear magnetic moments, without the $Z^2$ enhancement, are subdominant and neglected \cite{Brdar:2020quo}. The active neutrino masses are negligibly small comparing to the sterile neutrino mass $m_N$ we are considering, the neutrino energy $E_{\nu}$ and the nuclear recoil energy $T_n$ involved. As only protons contribute to this process, the enhancement factor only depends on $Z$. Note that this cross section can return to the scattering form of active neutrino with magnetic moment when $m_N=0$ is taken. The minimum neutrino energy to produce a sterile neutrino with mass $m_N$ in neutrino-nucleus upscattering is 
	\begin{equation}
	E_{{\rm min}}=\frac{m_N^2+2MT_n}{2\left[\sqrt{T_n(T_n+2M)}-T_n\right]}
	\label{Emin_ndp} \,.
	\end{equation}
Neutrino sources with higher neutrino energy can produce sterile neutrinos with larger $m_N$. Therefore, direct detection experiments based on SNS like COHERENT with energy as large as $\unit[50]{MeV}$ can probe parameter spaces for heavier sterile neutrino mass than those based on solar or reactor neutrinos.

The cross section of neutrino-electron scattering via the dipole portal can be written similarly as~\cite{Brdar:2020quo}
	\begin{equation}
 \begin{aligned}
	\frac{\dif \sigma^{\nu-e}_{\rm NDP}}{\dif T_e}&=d^2 \alpha_{\rm EM}\bigg[-\frac{m_N^2}{2 E_{\nu} m_eT_e}\left(1-\frac{m_e-T_e}{2E_{\nu}}\right)\\&-\frac{m_N^4(m_e-T_e)}{8E_{\nu}^{2}m_e^2T_e^2}+ \frac{1}{T_e}-\frac{1}{E_{\nu}}\bigg]\,,
  \end{aligned}
	\label{eq:NDPxser} 
	\end{equation}
The minimum neutrino energy to produce a massive sterile neutrino in neutrino-electron upscattering is 
	\begin{equation}
	E_{{\rm min}}=\frac{m_N^2+2m_eT_e}{2\left[\sqrt{T_e(T_e+2m_e)}-T_e\right]}
	\label{Emin_ndp_er} \,.
	\end{equation}
It should be noted that this minimum neutrino energy of neutrino source can be larger than that in the $\nu-N$ scattering case (Eq.~\ref{Emin_ndp}) to produce $N$ with mass $m_N$, since the target electron is much lighter than the nucleus and the recoiled electron will take away a considerable amount of the deposited energy ~\cite{Li:2022bqr}. 
\section{Expected Event Spectra}
\label{sec:spectra}
\begin{figure}
		\centering
		\includegraphics[scale=0.45]{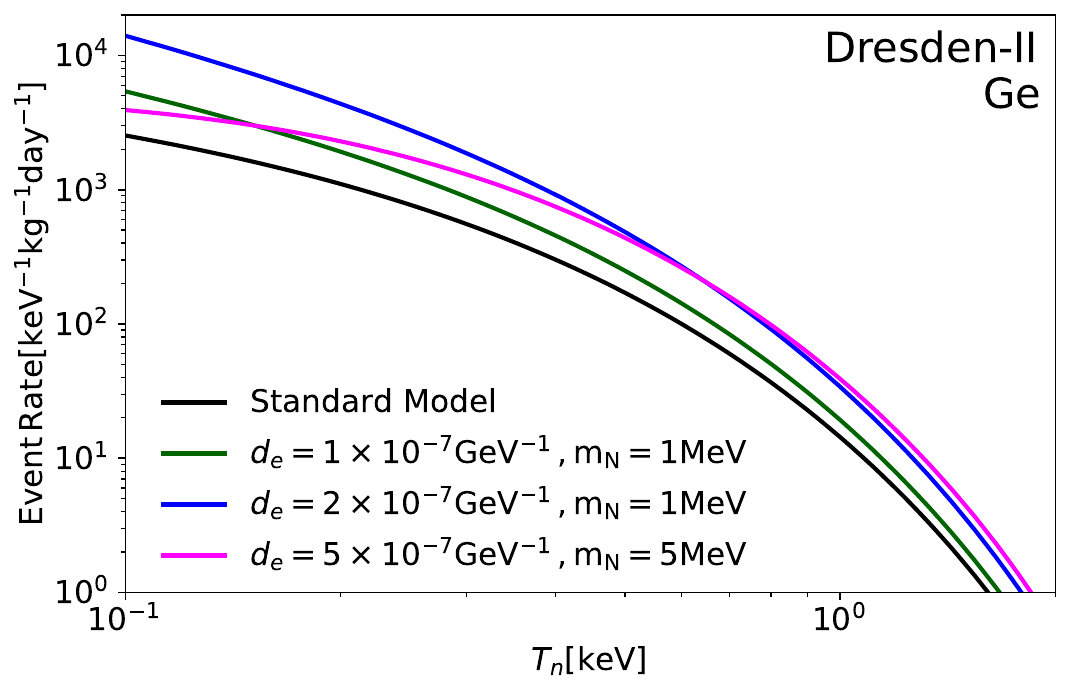}
		\includegraphics[scale=0.45]{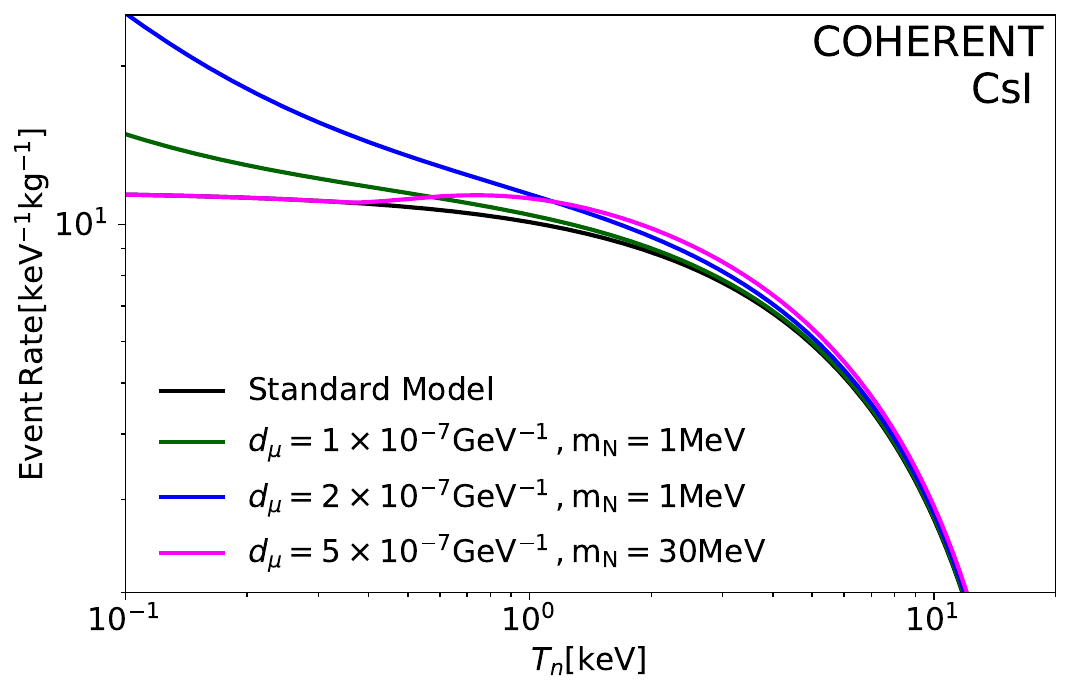}
		\includegraphics[scale=0.45]{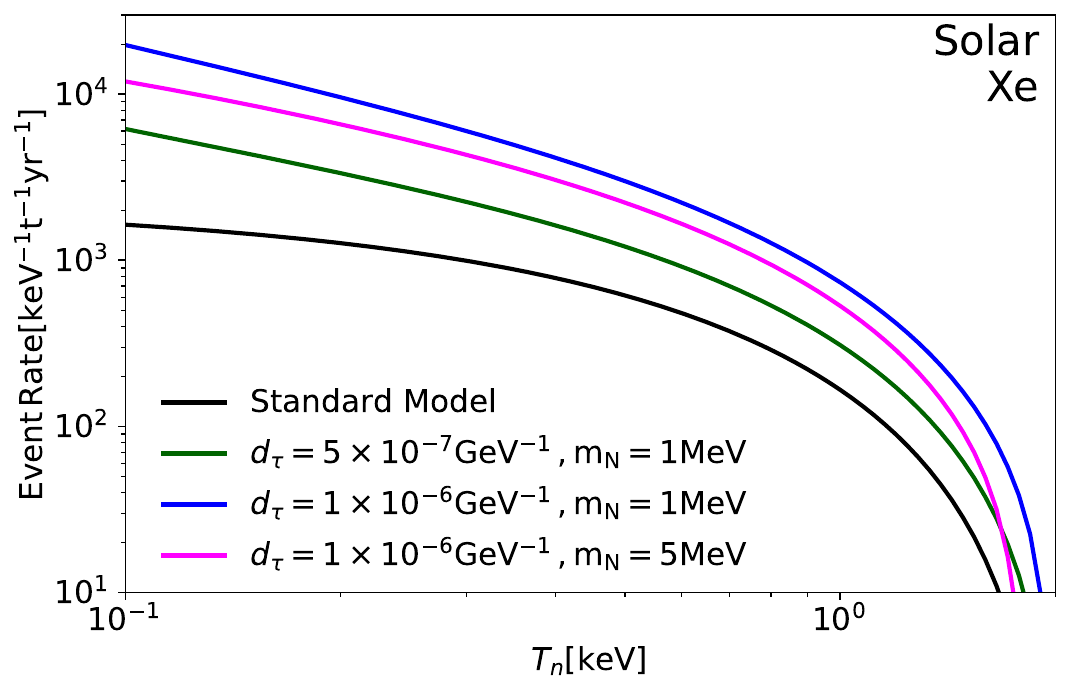}
		\caption{Event rate of CE$\nu$NS in germanium, CsI and xenon detectors with the corresponding sources of Dresden-II reactor, COHERENT SNS and solar neutrino, respectively, from top to bottom. The black lines show the event rate in the SM and the blue, green and purple lines show the event rate in the presence of $N$ for different benchmarks in order to illustrate the impact of $d$ and $m_N$ on the event spectra.}
		\label{fig:spec}
	\end{figure}
To probe $d_\alpha$ of a specific flavor and the $m_N$ in the range $\unit[0.1]{MeV}-\unit[50]{MeV}$, we consider experiments targeting on a large neutrino flux of flavor $\alpha$, which is neutrinos from reactor, SNS and the Sun for $\alpha = e$, $\alpha = \mu$ and $\alpha = \tau$, respectively. 
For $\alpha = e$, we analyze the large $\bar{\nu}_e$ flux from Dresden-II reactor based on the data collected by a 3 kg ultra low noise germanium detector in an exposure of 96.4 days~\cite{Colaresi:2022obx}. The reactor $\bar{\nu}_e$ spectra are obtained from the weighted average of $\bar{\nu}_e$ fluxes from isotopes $^{235}$U,  $^{238}$U,  $^{239}$Pu and $^{241}$Pu, which are the main contribution to the fission in the reactor. We will employ the $\bar{\nu}_e$ fluxes predicted from Ref.~\cite{Mueller:2011nm} to construct the expected $\bar{\nu}_e$ spectra above 2 MeV and for the low energy part we rely on the $\bar{\nu}_e$ fluxes predicted from Ref.~\cite{PhysRevD.39.3378}. 

In the CE$\nu$NS process, part of the nuclear recoil energy $T_{n}$ will be transformed into the electron-equivalent recoil energy $T_e$ which is described as the quenching effect $T_{e}(T_{n})$. The theoretical CE$\nu$NS event number in each bin of the kinetic energy $T_e$ of electron recoil can be calculated as 
	\begin{equation}
		\begin{aligned}
			N_{i}^{\rm{CE}\nu\rm{NS}}(\mathcal{N})=&N \dif t
			\int_{T_{e}^i}^{T_{e}^{i+1}}\dif T_e\int_{T_{n}^{\prime min}}^{T_{n}^{\prime max}}\dif T_{n}R(T_{e},T_{e}^{\prime}(T_{n}))\\
			\times &\int_{E_{min}}^{E_{max}}\dif E \frac{\dif N_{\bar{\nu}_e}}{\dif E }(E)\frac{\dif\sigma^{{\bar{\nu}_e}-\mathcal{N}}}{\dif T_{n}}(E, T_{n})\,,
		\end{aligned}
  \label{eq:cenns}
	\end{equation}
 where $\mathcal{N}$ is the different isotopes of germanium and $N=2.43\times10^{25}$ is the number of germanium atoms in the detector. The energy resolution function $R(T_{e}^{\prime},T_{e}(T_{n}))$ can be described as a truncated Gaussian \cite{AtzoriCorona:2022qrf}. The minimum ionization energy $T_{n}$ of germanium is 2.96 eV. The total CE$\nu$NS event number is integrating over contributions from different isotopes. 

The theoretical E$\nu$ES event number in each bin of the kinetic energy $T_e$ of electron recoil can be expressed in a similar way without the quenching effect as
	\begin{equation}
		\begin{aligned}
			N_{i}^{\rm{E}\nu\rm{ES}}(\mathcal{N})=&N_e \dif t
			\int_{T_{e}^i}^{T_{e}^{i+1}}\dif T_e\int_{T_{e}^{\prime min}}^{T_{e}^{\prime max}}\dif T_{e}^{\prime}R(T_{e},T_{e}^{\prime})\\
			\times &\int_{E_{min}}^{E_{max}}\dif E \frac{\dif N_{\bar{\nu}_e}}{\dif E }(E)\frac{\dif\sigma^{{\bar{\nu}_e}-e}}{\dif T_{e}^{\prime}}(E, T_{e}^{\prime})\,,
		\end{aligned}
  \label{eq:enes}
	\end{equation}	
	with the electron number in the target germanium $N_e({\rm Ge})=ZN$ where $Z=32$ is the atomic number of germanium.

For $\alpha = \mu$, we consider the full dataset collected by CsI[Na] detector\footnote{Given that a smaller number of events can be detected and the background is much larger~\cite{COHERENT:2021xmm,COHERENT:2020ybo}, we have neglected the Ar dataset from COHERENT in our analysis.} released by the COHERENT collaboration \cite{COHERENT:2021xmm, COHERENT:2018imc}. The CsI data contains complete information on the energy and arrival time of the observed events. The neutrino fluxes from SNS include a monochromatic component of $\nu_\mu$ flux from pion decay at rest $\pi^+\rightarrow\mu^++\nu_\mu$ and two components of $\bar{\nu}_{\mu}$ and $\nu_e$ that are produced in the subsequent muon decay $\mu^+\rightarrow e^++\bar{\nu}_{\mu}+\nu_e$. 
The maximum energy of $\bar{\nu}_\mu$ and $\nu_e$ fluxes is about half of the muon mass $E_{\rm max}=m_\mu/2\simeq 52.8\,\rm{MeV}$.
Since the $\nu_\mu$ flux is from the pion decay, earlier than the production of $\bar{\nu}_\mu$ and $\nu_e$ fluxes from muon decay, its arrival is within about 1 $\mu s$ after the beam-on trigger while the arrival of the other two fluxes is delayed with a time interval of up to 10 $\mu s$. Therefore, the time information is crucial to distinguish signals induced by the $\nu_\mu$ fluxes from those induced by other fluxes. As SNS also provides large neutrino flux of the $e$ flavor, CaI data can be employed to probe $d_e$. The theoretical CE$\nu$NS and E$\nu$ES events can be calculated similarly as that in \eq{cenns} and \eq{enes}, respectively, but with the quenching factor and the detector geometry replaced accordingly.

For $\alpha = \tau$, we consider the solar $^8$B neutrino flux which is produced from the fusion process inside the sun and including neutrinos of all three flavors after oscillation. The total neutrino-nucleus scattering events $N^{\nu-\mathcal{N}}_{{\rm NDP}}(\mathcal{N})$ for different isotopes $\mathcal{N}$ in the presence of sterile neutrino are given by
   	\begin{eqnarray}
   	N^{\nu-\mathcal{N}}_{{\rm NDP}}(\mathcal{N})&=& \frac{\epsilon}{M} \int_{T_{\rm min}}^{T_{\rm max}} \dif{T} \int_{E_{\rm min}}^{E_{\rm max}} \dif{E_{\nu}} \notag\\ &&\Phi (E_{\nu_\tau}) \eta\left(E_{\nu_\tau} \right)\frac{\dif{\sigma}_{{\rm NDP}}^{\nu-\mathcal{N}}}{\dif{T_n}}\,.
   	\end{eqnarray}
   where $\epsilon$ is the detector exposure, $M=131.293\rm{g/mol}$ is the isotope averaged mole mass of Xenon and $\Phi (E_{\nu_\tau})$ is the $^8$B neutrino flux. $\eta\left(E_{\nu_\tau} \right)$ is the energy dependent detector efficiencies for PandaX-4T \cite{PandaX:2022aac} and XENON1T \cite{XENON:2020gfr}, which is smaller than 1\% when the nuclear recoil energy $T_n\lesssim\unit[]{keV}$.

In \fig{spec} we present the expected event spectra of the CE$\nu$NS process, which are the main contributing processes to the signals within the mass range of $N$ considered in this study.  The spectra are presented for various experiments with the corresponding (anti)neutrino sources. Since for \fig{spec}, we aim at illustrate the impact of $d_\alpha$ on the event spectra, we have set the efficiency $\eta(E_\nu) = 1$, the quenching effect $T_e(T_{n}) = T_{n}$ and the resolution of the experiments $R(T_e, T'_e) = 1$. We show the event spectra in the SM and in the presence of the $N$. From top to bottom, results are shown for Dresden-II reactor antineutrino source with germanium detector, COHERENT experiment with CsI detector and solar neutrino source with xenon detector. In \fig{spec} we show the relevant spectra including the flavor-specific transition neutrino magnetic moment $d_\alpha$ of two different values for $m_N = \unit[1]{MeV}$, where the flavor of $\alpha$ and the value of $d_\alpha$ is indicated in the legend of each plot.
For these light $m_N$ cases, the up-scattering events are significantly increased at low recoil energies, which will be the main contribution to the improvements on probing the transitional magnetic moments. We also present the spectra for a heavier sterile neutrino case with $m_N = \unit[5]{MeV}$ and we observe that the event rate is suppressed since higher energy is required for the incident neutrino to upscatter to a heavier sterile neutrino. Also, it should be noted that the spectra on CsI detector with SNS source can expand to a higher recoil energy of more than 10 keV for a neutrino source with energy as high as 52.8 MeV. The recoil energy range for solar and reactor source is similar, but the reactor antineutrino flux is several orders of magnitude larger and is capable of induce recoil events more efficiently in the detector.
\begin{figure}[!htbp]
		\centering
		\includegraphics[scale=0.14]{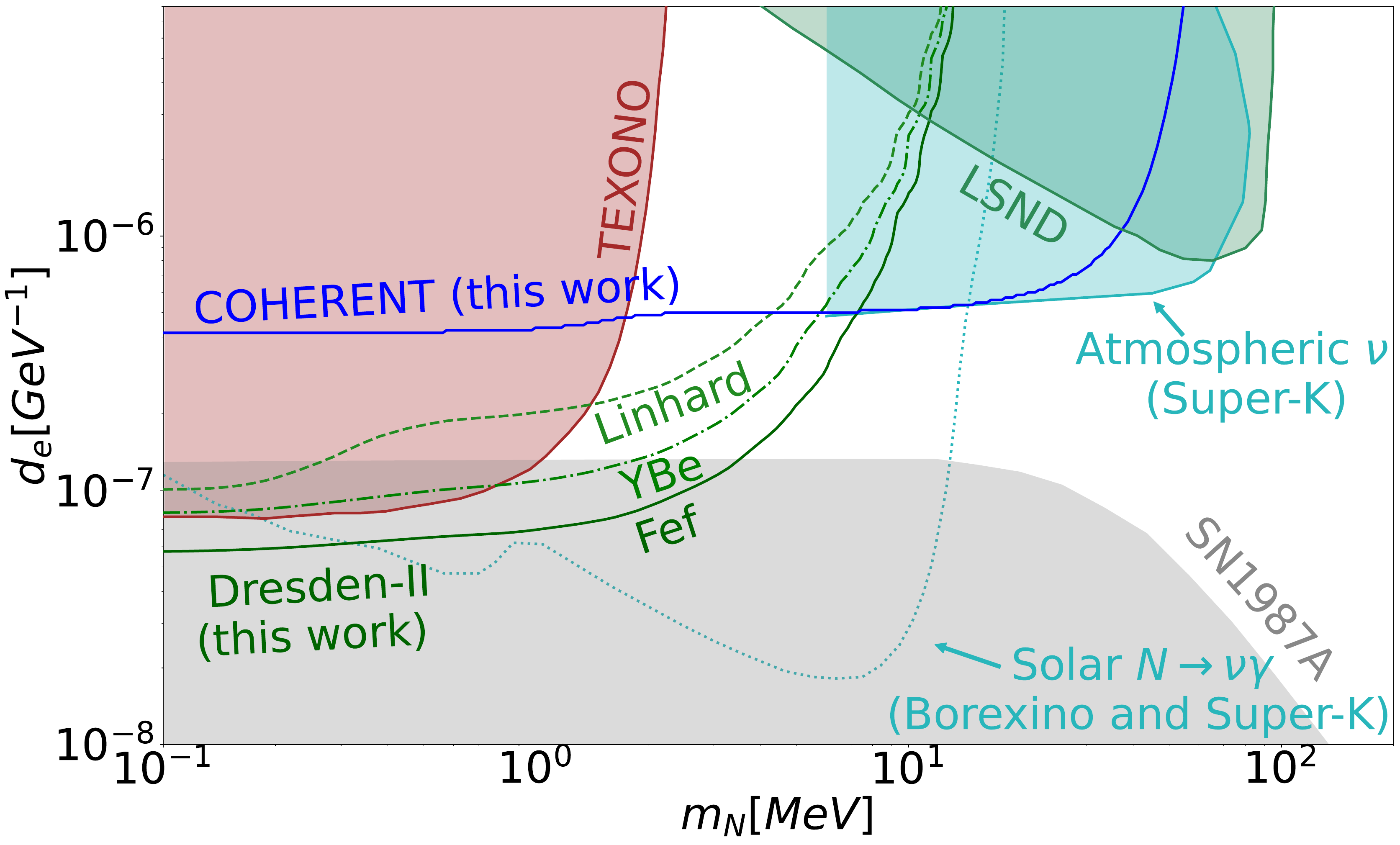}
		\includegraphics[scale=0.14]{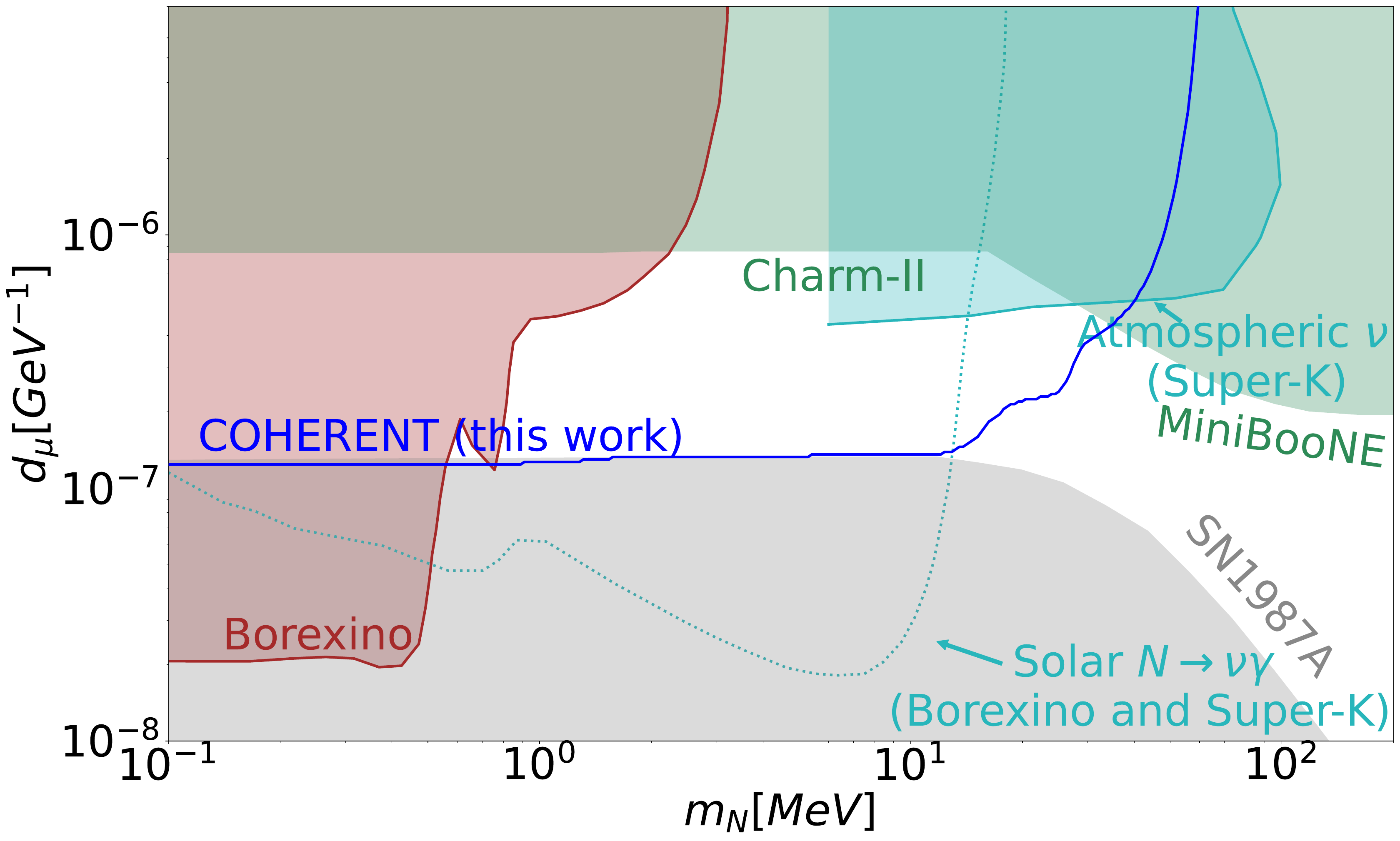}
        \includegraphics[scale=0.14]{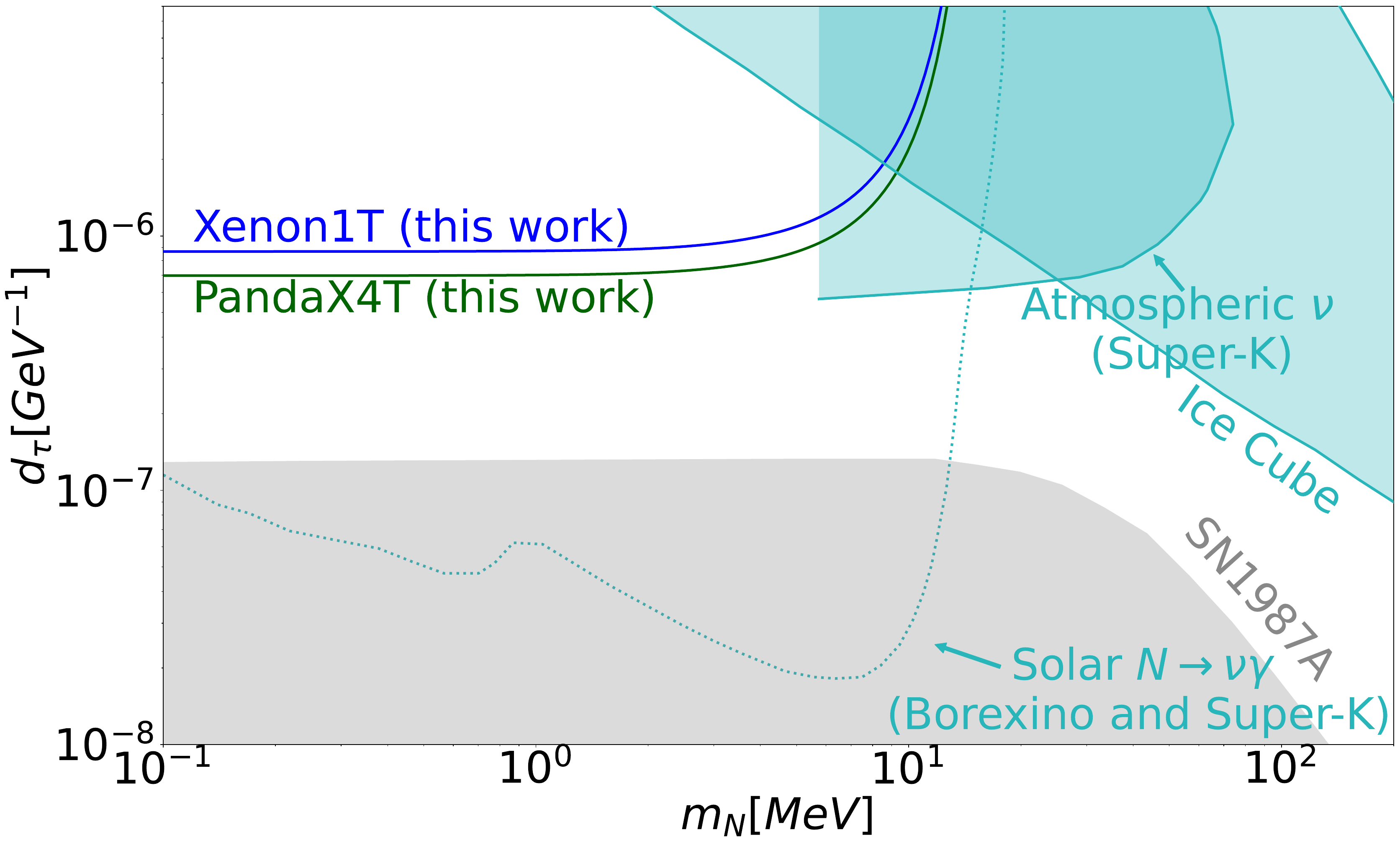}
		\caption{Numerical sensitivities for sterile neutrino the 90$\%$ C.L. constraints on the neutrino transition magnetic moments for different flavors and the corresponding sterile neutrino mass. From top to bottom results are shown for constraints on the neutrino transition magnetic moments of flavor $e$, $\mu$ and $\tau$. For the Dresden-II analysis, we show the results with different quenching models of Linhard theory, YBe and Fef. We also include constraints from other experiments including SN1987A~\cite{Magill:2018jla}, TEXONO~\cite{PhysRevD.81.072001}, LSND~\cite{Magill:2018jla}, Atomspheric $\nu$ (Super-K)~\cite{Gustafson:2022rsz} which looks for atomspheric neutrino upscattering to $N$ with subsequent $N\to \nu\gamma$ decay,  Borexino~\cite{Brdar:2020quo,Redchuk:2020hjv}, Solar $N\to \nu\gamma$ (Borexino and Super-K) \cite{Plestid:2020vqf} for solar neutrino upscattering to $N$ with subsequent $N$ decays, MiniBooNE~\cite{Shoemaker:2018vii,Magill:2018jla,MiniBooNE:2007uho}, Charm-II~\cite{Shoemaker:2018vii,CHARM-II:1989srx} and IceCube~\cite{Coloma:2017ppo} for comparison. All limits are for the one flavor transitional magnetic moment, except for the SN case where flavor universal couplings are used. }\label{fig:sens}
\end{figure}
\section{Constraints}
\label{sec:limts}
We illustrate the sensitivity to probe the transition neutrino magnetic moments of different flavor.  For Dresden-II to probe $d_e$, predictions of quenching effect from the Linhard theory is in good agreement with the experimental data for $T_e$ around keV range~\cite{Bonhomme:2022lcz,Li:2022xbv}. However, in a recent experiment, an unexpected enhancement on quenching factor for $T_e$ in the sub keV region is observed~\cite{PhysRevD.103.122003}, which is in great tension with the Linhard theory. Therefore, in this work we will also consider two more commonly used quenching models from photon-neutron source measurements (YBe)~\cite{PhysRevD.103.122003} and iron-filtered monochromatic neutrons (Fef)~\cite{PhysRevD.103.122003} other than the Linhard at $k=0.157$~\cite{osti_4701226}. For CsI data from COHERENT experiment, only the Fef quenching model is employed. In order to include the time information of the CsI data of COHERENT, we calculate the theoretical event number $N_{ij}^{\rm{pred}}$ at the energy bin $i$ and time bin $j$ based on the neutrino flux with both energy and time dependences. We then employ the Poissonian least-squares function to analyze the data from Dresden-II and COHERENT following \cite{AtzoriCorona:2022qrf}. 

In \fig{sens} we present the 90$\%$ C.L. constraints on $d_\alpha$ of a specific flavor as a function of sterile neutrino mass $m_N$. The signals from CE$\nu$NS process dominate the contribution to the constraints for heavy sterile neutrino ($m_N \gtrsim \unit[]{MeV}$) due to kinematic constraints, while at low sterile neutrino mass the contribution from E$\nu$ES prevails for a lower energy threshold on the incident neutrinos~\cite{Li:2022bqr}. Constraints from other experiments are also shown for comparison, among which the constraints from SN1987A~\cite{Magill:2018jla} are considering flavor universal couplings to $N$ via the dipole portal. 
For $d_e$, the constraints obtained are shown in the top plot of \fig{sens}. For the Dresden-II analysis, we show the results with different quenching models of Linhard theory, YBe and Fef described above. The Ybe and Fef quenching model enhance the response of germanium detector in sub-keV energy range and significantly improve the constraints on $d_e$. The dense reactor antineutrino flux provide stringent constraints at low mass range that exceed the existing limits obtained at Borexino and Super-K by searching for solar neutrino upscattering to $N$ with subsequent $N\to \nu\gamma$ (cyan dashed line) \cite{Gustafson:2022rsz}, and at TEXONO~\cite{PhysRevD.81.072001} by measuring neutrino-electron scattering.
Among the upscattering measurements, the results from Dresden-II is competitive between 0.1 MeV and 10 MeV. 
The constraints from COHERENT can expand to a mass of about 50 MeV due to a high energy source, which are competitive as that from LSND~\cite{Magill:2018jla} and the non-observation of atmospheric neutrino upscattering to N with subsequent $N$ decays at Super-K for $e$ flavor (cyan shaded region)~\cite{Gustafson:2022rsz}. 

The $\mu$ flavor neutrino flux from the SNS source in COHERENT is more intense than that of $e$ flavor and the COHERENT CsI data as a result can provide a better constraints on the dipole operator of $\mu$ flavor. 
Note that the full dataset we employ includes about 3 times more statistics than that in previous works \cite{Miranda:2021kre, Bolton:2021pey}. The systematic uncertainties are also reduced by a factor of 2 to 3. Furthermore, we adopt a two-dimensional analysis method that includes not only the energy information, but also the time information. Contributions to upscattering induced by $\nu_\mu$ fluxes can now be separated from those induced by $\bar{\nu}_\mu$ and $\bar{\nu}_e$ fluxes based on the arrival time. As a result, our limits for the COHERENT experiment improve over previous ones \cite{Miranda:2021kre, Bolton:2021pey} significantly, by roughly one order of magnitude. Though in the mass range lower than $\unit[1]{MeV}$, the constraint from COHERENT CsI data is not as good as the results from solar neutrino E$\nu$ES observation at Borexino~\cite{Brdar:2020quo,Redchuk:2020hjv}, it provides better constraints for higher masses. We also show the constraints from accelerator experiments including MiniBooNE~\cite{Shoemaker:2018vii,Magill:2018jla,MiniBooNE:2007uho} and Charm-II~\cite{Shoemaker:2018vii,CHARM-II:1989srx} and the non-observation of atmospheric neutrino upscattering to N with subsequent $N$ decays at Super-K for $\mu$ flavor (cyan shaded region)~\cite{Gustafson:2022rsz}. It should be noted that the constraints on $d_\mu$ from COHERENT CsI data exclude a unique region in the parameter space between about $\unit[10]{MeV}$ and $\unit[40]{MeV}$ and the monochromatic peak of $\nu_\mu$ from the rest pion decay further lower the border of exclusion region at $m_N\simeq\unit[30]{MeV}$.

For $\alpha = \tau$, PandaX-4T and Xenon1T reported the 90$\%$ C.L. upper limit of $^8$B solar neutrino flux to be $\Phi_{90\%}\simeq\unit[9.0\times10^6]{cm^{-2}s^{-2}}$ ~\cite{PandaX:2022aac} and $\Phi_{90\%}\simeq\unit[1.4\times10^7]{cm^{-2}s^{-2}}$  ~\cite{XENON:2020gfr}, respectively. We utilize these two upper limits to provide constraints on the sterile neutrino via dipole portal. Since the $^8$B neutrino flux from the B16 Standard Solar Model (SSM)~\cite{Bergstrom:2016cbh} determination is $\Phi_{\rm SSM}\simeq 5.16\times10^6\,\,\rm{cm}^{-2}\rm{s}^{-2}$, the constraints can be expressed as 
   	\begin{equation}
   		\frac{\left\langle N^{\nu-\mathcal{N}}_{{\rm SM+NDP}}(\mathcal{N}) \right\rangle }{\left\langle N^{\nu-\mathcal{N}}_{{\rm SM}}(\mathcal{N}) \right\rangle }\lesssim \frac{\Phi_{90\%}}{\Phi_{\rm SSM}}
   	\end{equation}
where $\mathcal{N}$ is different isotopes of xenon in the detector and $\left\langle \cdot \right\rangle $ denotes the isotopic average performed on the event number based on the natural abundance of xenon. The subscript indicate the event number predicted from neutrino magnetic dipole portal and the SM scenario. With this, we show the exclusion limit for Xenon1T and PandaX in the bottom plot of \fig{sens}, which for Xenon1T is a factor of 3 better than the limit on $d_\mu$ obtained considering nuclear recoil for solar neutrinos in Ref.~\cite{Brdar:2020quo}.
This improvement is primarily due to the significantly lower nuclear recoil energy threshold of 0.5 keV achieved in Refs.\cite{XENON:2020gfr, PandaX:2022aac}, compared to the 4.5 keV threshold used in Ref.~\cite{Brdar:2020quo}. 
As can be seen in \fig{sens}, this region can also be fully probed by Borexino and Super-K by searching for solar neutrino of $\tau$ flavor upscattering to $N$ with subsequent $N$ decays (cyan dashed line)~\cite{Gustafson:2022rsz}, and the cosmic neutrino observation at IceCube~\cite{Coloma:2017ppo} at higher $m_N$. The results from Xenon1T and PandaX-4T solar neutrino observation provide independent constraints at mass lower than 10 MeV.

\section{Conclusions}
\label{sec:conclusion}
In this work, we explore the potential of the recently collected CE$\nu$NS and E$\nu$ES data at the Dresden-II and COHERENT experiments in probing sterile neutrinos via the magnetic dipole portal. Given the large electron anti-neutrino flux from reactors, we find that the current Dresden-II data provide leading probes of the $e$-flavor dipole operator for $m_N \lesssim \unit[0.5]{MeV}$ in the case where Fef quenching factor is used. Future research should try to reconcile  predictions between various quenching models and also between predictions and data. With the large $\nu_\mu$ flux from spallation neutron source, we demonstrate that utilizing the timing information in the full dataset collected by CsI detector from COHERENT experiment, a unique parameter region can be probed for heavier $m_N$ up to $\unit[40]{MeV}$ in the $\mu$-flavor dipole operator case.
For $\tau$ flavor transitional dipole portal, measurements of solar neutrino fluxes at PandaX-4T and Xenon1T probes a parameter region that is covered by the Borexino and Super-K experiments from the non-observation of the $N$ decays with $N$ produced via solar neutrino upscattering. It is also valuable to explore strategies searching for $N$ in the large uncovered parameter space for $m_N \gtrsim \unit[40]{MeV}$ above the cooling band, which will be left for future work.
\begin{acknowledgments}
We would like to thank Vedran Brdar, Fei Gao, Kun-Feng Lyu for inspiring discussions.
Y.-Y. L is supported by the NSF of China through Grant No. 12305107, 12247103. Y.F. L and S.Y. X are supported in part by National Natural Science Foundation of China under Grant Nos.~12075255, 12075254 and 11835013.
\end{acknowledgments}

\bibliography{main}

\end{document}